\documentclass[12pt]{article}
\usepackage[dvipdfmx]{graphicx}
\begin{document}
\title{Pointer States and Decoherence in Quantum Unitarity: \\
Energy Conservation for Weak Interaction Model}

\author{Kentaro Urasaki\\
Suginami, Tokyo, Japan, urasaki@ynu.ac.jp}
\date{}%

\maketitle

\abstract{
The purpose of the present paper is 
to derive the pointer states of a macro-object
using a simple perturbation method.
We study the model Hamiltonian involving 
the weak interaction between the center of mass and 
its environment.

The main conclusion is that 
the pointer states emerge as the small part of 
the total state vector
when the interaction Hamiltonian is not dominant: 
the degrees of freedom being at the center of interaction can be localized. 
In the weak interaction case, 
if the disappearance of the energy contribution from the off-diagonal terms 
is caused by the decoherence, it threatens the conservation law.
In this study, however, the energy conservation reasonably holds, 
since the non-classical (i.e., normal quantum) states give main contribution to it. 

The emergence of the pointer states also implies the special initial state 
in which the localized state makes the deep well of the potential. 
In the result, the unitary time evolution of a pointer state is very different 
from that of the majority of the original state vector of the total system.
}

\section{Introduction}
The degrees of freedom showing classical behavior are only the small part of the enormous number of that consisting of a marco-object. 
These states, however, are the frame of our daily experience.
Therefore understanding of the quantum decoherence means 
shedding light on
the origin of the classical existence 
with the use of the established concept of the quantum unitary evolution\cite{Joos2003, Zurek2003}.

The essence of the quantum decoherence is the interaction between 
a macro-system and its environment.
For this purpose, models consisting 
only of the interaction Hamiltonian are often used. 
Namely the time evolution by the system's self-Hamiltonian are ignored. 
These simple models are effective to understand 
the essencial mechanism of the quantum decoherence.
However, in exchange for it, 
these are not suited for clarifying 
the competition between the decoherence by 
the interaction and the coherence by the 
self-evolution. 
In such a situation, 
the previous work\cite{Paz1999} 
reasonably shows that 
the pointer states do not appear as the eigenstate of 
the interaction Hamiltonian
when the system-environment 
interaction is weak. 

In the present study, however, we find out 
that the pointer states can appear even for the weak interaction if the concerned degrees of freedom is at the {\it center of entanglement}: 
it is also important to notice that 
the pointer states constitute only the small part of 
the total state vector. 
For example, the center of mass (COM) position 
effectively influences on all of other degrees of freedom so that 
the state of COM is naturally entangled with these states.

\section{Quantum Decoherence and Unitary Evolution}

\subsection{Environment-induced Decoherence}

In usual, the simplest explanation of the quantum decoherence is derived 
under the assumption that the interaction Hamiltonian is dominant in the time evolution\cite{Joos2003}.
It will be still important to take a glance at such the simplest case 
in order to advance toward 
more realistic cases.
%the essence of the general case.

Let us start from the following initial (product) state of the total system:
\begin{eqnarray}\label{vNform-1}
|\Phi(0)\rangle=(c_1|\phi_1\rangle+c_2|\phi_2\rangle)|\varepsilon\rangle.
\end{eqnarray}
The states, $|\phi_1\rangle, |\phi_2\rangle$, are the system's eigenstates of 
the interaction Hamiltonian, $\hat{h}_I$, 
while the state, $|\varepsilon\rangle$, are the environment's initial state.
If we assume an appropriate interaction, 
it evolves into 
\begin{eqnarray}\label{vNform-2}
|\Phi(t)\rangle=c_1|\phi_1(t)\rangle|\varepsilon_1(t)\rangle+c_2|\phi_2(t)\rangle
|\varepsilon_2(t)\rangle,
\end{eqnarray}
where two states, $|\phi_1(t)\rangle, |\phi_2(t)\rangle$, are stable against the interaction and called as the pointer states.
For the operator $\hat{Q}$ acting only on the system $\phi$, 
its expectation value is expressed as, 
\begin{eqnarray}
\langle \hat{Q} \rangle=\!\!\!\!&&|c_1|^2\langle\phi_1(t)|\hat{Q}|\phi_1(t)\rangle+|c_2|^2\langle\phi_2(t)|\hat{Q}|\phi_2(t)\rangle\\
&&+c^\ast_1 c_2\langle\phi_1(t)|\hat{Q}|\phi_2(t)\rangle\langle\varepsilon_1(t)|\varepsilon_2(t)\rangle
+c^\ast_2 c_1\langle\phi_2(t)|\hat{Q}|\phi_1(t)\rangle\langle\varepsilon_2(t)|\varepsilon_1(t)\rangle.
\end{eqnarray}
If the state of the environment develop into the corresponding orthogonal state
after the interaction, we can say, 
\begin{eqnarray}
\langle\varepsilon_1(t)|\varepsilon_2(t)\rangle\to 0.
\end{eqnarray}
In this case, the coherent (off-diagonal) terms in the above equation can be neglected. 
Since the above equation becomes similar to that of statistical mixture of the events, 
we can also say the {\it approximate} mixture is obtained. 
This process is called decoherence.\footnote{
We also find out the same result  
introducing the reduced density operator for the macroscopic system as,
\begin{eqnarray}
\rho_\phi:={\rm Tr}_\varepsilon|\Phi(t)\rangle\langle\Phi(t)|
\to p_1(t)|\phi_1\rangle\langle\phi_1|+p_2(t)|\phi_2\rangle\langle\phi_2|, 
\end{eqnarray}
where the density matrix for mixed states in the privileged Schmidt basis, $|\phi_1\rangle, |\phi_2\rangle$, appears. }
These non-obvious results from the unitary evolution surely explain some part of the classicality\cite{Joos2003}.

\bigskip

In this case, in eq. (\ref{vNform-2}), the superposition of all the branch states is assumed to be identified as the total state vector: 
the unitary evolution by the system's Hamiltonian
is assumed not to mix these states.
Such the assumption can be justified when the system- and 
the interaction Hamiltonian are approximately commutative. 
(For another example, we can find the measurement model 
by von Neumann in \cite{Neumann1932}.)
It, however, is necessary to reexamin this understanding that the total state vector 
is completely devided into branch states, when the interaction Hamiltonian
is not dominant.
For example, this simple scenario gets into difficulty if we consider the energy conservation for the non-commutative case.

\subsection{Viewpoint of Energy Conservation}
In free space, let us consider the long wave limit of a plane wave state, 
$|\phi_{GS}(t)\rangle$, which is the ground state of the Hamiltonian. 
The energy of this system is a constant value: 
\begin{eqnarray}
\langle\phi_{GS}(t)|\hat{h}_0|\phi_{GS}(t)\rangle=E_0.
\end{eqnarray}
We expand the plane wave state with the localized states at $t=0$: 
$|\phi_{GS}(0)\rangle=\sum_{\bf R}\alpha_{\bf R}|\phi({\bf R})\rangle$.
\footnote{For our purpose, 
it is enough that the `localized state' is understood as a very narrow wave packet.}
So that the left hand side of the above equation becomes, 
\begin{eqnarray}
{\rm lhs}&&=
\sum_{{\bf R}, {\bf R}^\prime}\alpha^\ast_{\bf R}\alpha_{{\bf R}^\prime}\langle\phi({\bf R}, t)|\hat{h}_0|\phi({\bf R}^\prime, t)\rangle\\
&&
%\hspace{-2cm}
=\sum_{\bf R}|\alpha_{\bf R}|^2\langle\phi({\bf R}, t)|\hat{h}_0|\phi({\bf R}, t)\rangle
+\sum_{{\bf R}\ne{\bf R}^\prime}\alpha^\ast_{\bf R}\alpha_{{\bf R}^\prime}\langle\phi({\bf R}, t)|\hat{h}_0|\phi({\bf R}^\prime, t)\rangle.
\end{eqnarray}
For easy to see, the on- and off-diagonal terms are separated in the second equation. 
The state, $|\phi({\bf R}, t)\rangle:=e^{-i\hat{h}_0t/\hbar}|\phi({\bf R})\rangle$, is localized at potision ${\bf R}$ at
$t=0$, and spreads over due to the Hamiltonian, $\hat{h}_0$.

Since all of these states have much higher energy compared with the ground state, the whole contribution from the diagonal elements is also much higher energy than $E_0$:  
\begin{eqnarray}
\sum_{\bf R}|\alpha_{\bf R}|^2\langle\phi({\bf R}, t)|\hat{h}_0|\phi({\bf R}, t)\rangle\gg E_0.
\end{eqnarray}
(We notice, $\displaystyle \sum_{\bf R}|\alpha_{\bf R}|^2=1$.)
Naturally the off-diagonal terms give negative contribution to the total energy. 
These terms are necessary for the energy conservation.

\bigskip

Next, we assume that 
the decoherence caused by the weak interaction between a macro-object and its environment.

The plane wave states of the center-of-mass (COM) 
are the eigen states of the system's Hamiltonian $\hat{h}_0$. 
On the other hand, the localized states of the COM, $\{|{\bf R}\rangle\}$ are 
the eigen states of the interaction Hamiltonian, $\hat{h}_I$.
(Namely, we are interested particulary in the case  $\hat{h}_0$ and $\hat{h}_I$ are not commutative.)

We assume the weak interaction: 
the interaction energy is a macroscopic value but negligible compared to the system's energy, 
$\langle\hat{h}_0\rangle \gg\langle \hat{h}_I\rangle$.
If the decoherence scenario for the strong correlation is directly applicable to the present case with no revision, 
the unitary time evolution leads that the contribution to the energy from the off-diagonal terms disappears:
\begin{eqnarray}\label{self-energy}
\langle\Phi_{GS}(t)|\hat{h}_0|\Phi_{GS}(t)\rangle
&&=
\sum_{\bf R}|\alpha_{\bf R}|^2\langle\phi({\bf R}, t)|\hat{h}_0|\phi({\bf R}, t)\rangle\nonumber\\
&&+\sum_{{\bf R}\ne{\bf R}^\prime}\alpha^\ast_{\bf R}\alpha_{{\bf R}^\prime}\langle\phi({\bf R}, t)|\hat{h}_0|\phi({\bf R}^\prime, t)\rangle
\langle\varepsilon_{\bf R}(t)|\varepsilon_{{\bf R}^\prime}(t)\rangle\\
{\rm decoherence}&&\to \sum_{\bf R}|\alpha_{\bf R}|^2\langle\phi({\bf R}, t)|\hat{h}_0|\phi({\bf R}, t)\rangle\gg E_0?
\end{eqnarray}
Here, $|\Phi_{GS}(0)\rangle:=|\phi_{GS}(0)\rangle|\varepsilon(0)\rangle$.
We have used the diagonality among the states of the environment, 
 $\langle\varepsilon_{\bf R}(t)|\varepsilon_{{\bf R}^\prime}(t)\rangle\to \delta_{\bf RR^\prime}$, 
which is the result of the unitary time evolution. 

Of course, because of the interaction, the expectation value of the self-Hamiltonian, $\langle\Phi(t)|\hat{h}_0|\Phi(t)\rangle$,
is no longer a conserved quantity. 
In spite of this, it is important that, for the ground state of the COM, 
being without any kind of instability,  
the weak perturbation cannot make it drastically change 
by the unitary evolution.

Therefore the lhs of eq. (\ref{self-energy}) should be a almost constant value 
even when the weak interaction exists. 
{\bf Under the unitary time evolution, the effect of the weak peruturbation on the system's energy should be small even when the decoherence exists.}
This implies the logical inconsistency 
on the assumption that the scenario of the strong correlation 
can be directly applicable to the weak correlation case.

\bigskip

In this subsection, we have confirmed the inconsistency between 
the disappearance of all of the off-diagonal terms and the energy conservation in the weak interaction case.
Even for the weak interaction, 
the decoherence, however, can be found out as a small part of the whole state of 
the total system as seen below. 
In this case, in contrast to the strong interaction case, 
it is important that 
the all of the eigen states of the interaction Hamiltonian do not behave classical. 
The non-classical, i.e., usual quantum, states 
give the primary contribution to the energy, 
where the coherence caused by the self-Hamiltonian is dominant.

\section{Pointer States in the Perturbation Model}
For example, 
the fluctuation of short length scale in the external field 
compared to the size of a macro-object 
causes the effective interaction between the COM and other internal degrees of freedom exists. 
The weak interactions can be assumed to be independent each other (Fig. 1). 

In this section, firstly we find out the peculiarlity of such the interaction: 
the huge number of the environmental states suppresses 
the transition by the interaction. 
Although, for easy treatment, we investigate the two level system embeded in the environment below, 
the essence of the emergent properties will be typical.

\bigskip

\underline{\bf Model}\\

For easy treatment, we assume $\phi$ and each site in Fig. 1 are two states systems.
The environmental state is represented as $|\varepsilon\rangle=|\ \ \rangle_1|\ \ \rangle_2\cdots|\ \ \rangle_M$,
being expressed with $N=2^M$ basis.

\begin{figure}
\begin{center}
\includegraphics[width=7cm]{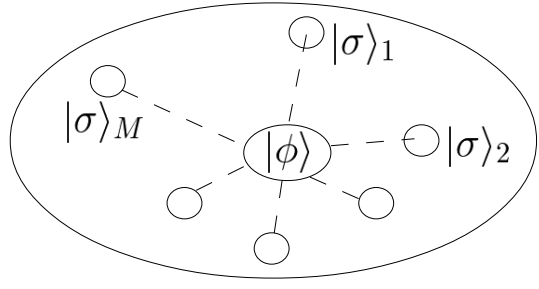}
\caption{The system and the environment (image).}
\end{center}
\end{figure}

\begin{figure}
\begin{center}
\includegraphics[width=3cm]{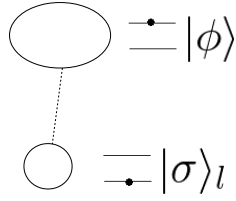}
\caption{The single interaction between the system and the l-th site (image).}
\end{center}
\end{figure}

We set the self-Hamiltonian as,  
\begin{eqnarray}
\hat{h}_{\phi}=E\ (|E_{+}\rangle\langle E_{+}|-|E_{-}\rangle\langle E_{-}|),\\
\hat{h}_l=\omega_l(|+\rangle\langle+|_l-|-\rangle\langle-|_l),
(l=1, \dots, M).
\end{eqnarray}
To study the case that the interaction Hamiltonian and the self-Hamiltonian are not commutative,
these energy eigenstates are assumed to be represented in the eigenstates of the interaction Hamiltonian, $|\phi_1\rangle$ and $|\uparrow\rangle_l$, as, 
\begin{eqnarray}
|E_\pm\rangle=\frac{1}{\sqrt{2}}(|\phi_1\rangle\pm|\phi_2\rangle),\\
|\pm\rangle_l=\frac{1}{\sqrt{2}}(|\uparrow\rangle_l\pm|\downarrow\rangle_l).
\end{eqnarray}
We define the self-evolution of these states:
\begin{eqnarray}
|\phi_1(t)\rangle:=&&e^{-i\hat{h}_\phi t}|\phi_1\rangle\\
=&&\frac{e^{-iE t}}{\sqrt{2}}|E_+\rangle+\frac{e^{iE t}}{\sqrt{2}}|E_-\rangle\\
=&&\cos E t |\phi_1\rangle-i\sin E t |\phi_2\rangle,\\
|\uparrow(t)\rangle_l:=&&e^{-i\hat{h}_l t}|\uparrow\rangle_l\\
=&&\cos \omega_l t |\uparrow\rangle_l-i\sin \omega_l t |\downarrow\rangle_l,
\end{eqnarray}
and so on (we use $\hbar=1$ in this subsection).

The total Hamiltonian is, 
\begin{eqnarray}
\hat{h}=\hat{h}_0+\hat{h}_I.
\end{eqnarray}
We here define the self-Hamiltonian as,
\begin{eqnarray}
\hat{h}_0:=\hat{h}_{\phi}\otimes\hat{1}_\varepsilon+\hat{1}_\phi\otimes\hat{h}_\varepsilon,
\end{eqnarray}
where, 
\begin{eqnarray}
\hat{h}_\varepsilon:=\sum_l\hat{h}_{l}\ \prod_{l^\prime\neq l}\otimes\hat{1}_{l^\prime}.
\end{eqnarray}
On the other hand, the total interaction is defined as, 
\begin{eqnarray}
\hat{h}_I:=\sum_l\hat{h}_{\phi, l}\ \prod_{l^\prime\neq l}\otimes\hat{1}_{l^\prime}.
\end{eqnarray}
The matrix elements of the interaction Hamiltonian are diagonal in $|\phi_1\rangle, |\phi_2\rangle, |\uparrow\rangle_l$ 
and $|\downarrow\rangle_l$,
\begin{eqnarray}
\hat{h}_{\phi, l}=\sum_{i=1,2}\sum_{\sigma=\uparrow, \downarrow}v_{i, \sigma}^l|\phi_i\rangle\langle\phi_i|\otimes|\sigma\rangle\langle\sigma|_l.
\end{eqnarray}

The case, $E=\omega_l=0$, (the self-enegies are zero) can be identified to the Zurek's strong interaction model\cite{Zurek1981}.

\vspace{1cm}
\underline{\bf Perturbation approach}\\

The initial states of the environment are labeled as,
\begin{eqnarray}
|\varepsilon_1(0)\rangle:=
|\uparrow\rangle_1 |\uparrow\rangle_2 |\uparrow\rangle_3 \cdots  |\uparrow\rangle_{M-1}|\uparrow\rangle_M, \nonumber\\
|\varepsilon_2(0)\rangle:=
|\downarrow\rangle_1 |\uparrow\rangle_2 |\uparrow\rangle_3 \cdots  |\uparrow\rangle_{M-1}|\uparrow\rangle_M, \nonumber\\
|\varepsilon_3(0)\rangle:=
|\uparrow\rangle_1 |\downarrow\rangle_2 |\uparrow\rangle_3 \cdots  |\uparrow\rangle_{M-1}|\uparrow\rangle_M, \nonumber\\
\cdots,\nonumber\\
|\varepsilon_{M+1}(0)\rangle:=
|\uparrow\rangle_1 |\uparrow\rangle_2 |\uparrow\rangle_3 \cdots  |\uparrow\rangle_{M-1}|\downarrow\rangle_M, \nonumber\\
\cdots, \nonumber\\
|\varepsilon_N(0)\rangle:=
|\downarrow\rangle_1 |\downarrow\rangle_2 |\downarrow\rangle_3 \cdots  |\downarrow\rangle_{M-1}|\downarrow\rangle_M.
\end{eqnarray}
The self-evolution is, 
\begin{eqnarray}
|\varepsilon_1(t)\rangle:=
e^{-i\hat{h}_\varepsilon t}|\uparrow\rangle_1 |\uparrow\rangle_2 \cdots |\uparrow\rangle_M
=\prod_{l=1}^{M} (\cos \omega_l t|\uparrow\rangle_l-i\sin \omega_l t|\downarrow\rangle_l),
\end{eqnarray}
and so on.
We advance toward the approximation. 
Substituting the expansion by non-perturbative states with the time-dependent coefficients, 
\begin{eqnarray}\label{expansion}
|\Phi(t)\rangle=\sum_{\nu^\prime=1}^N\sum_{j=1, 2}c_{j\nu^\prime}(t)|\phi_j(t)\rangle|\varepsilon_{\nu^\prime}(t)\rangle,
\end{eqnarray}
into the exact Schr\"odinger equation, $[i\hbar\partial_t-\hat{h}]|\Phi(t)\rangle=0,$ we obtain,
\begin{eqnarray}\label{expansion-Seq}
\sum_{\nu^\prime=1}^N\sum_{j=1,2}i\hbar\ \dot{c}_{j\nu^\prime}(t)|\phi_j(t)\rangle|\varepsilon_{\nu^\prime}(t)\rangle=\hat{h}_{I}\sum_{\nu^\prime=1}^N\sum_{j=1, 2}c_{j\nu^\prime}(t)|\phi_j(t)\rangle|\varepsilon_{\nu^\prime}(t)\rangle.
\end{eqnarray}
As already mentioned, the time dependence of the states, 
$|\phi_j(t)\rangle$ or $|\varepsilon_{\nu^\prime}(t)\rangle$, indicates the self-evolution.

\vspace{1cm}

Of course, the self-evolution does not change the orthogonality relation, 
\begin{eqnarray}
\langle\sigma(t)|\sigma^\prime(t)\rangle_{l^\prime}=\delta^{(l^\prime)}_{\sigma\sigma^\prime}, \\
\langle\varepsilon_{\nu}(t)|\varepsilon_{\nu^\prime}(t)\rangle=\delta_{\nu\nu^\prime},
\end{eqnarray}
where $\sigma=\uparrow, \downarrow$.
On the other hand, the matrix elements fluctuate around the initial values as, 
\begin{eqnarray}
\langle\varepsilon_\nu(t)|\hat{h}_I|\varepsilon_{\nu^\prime}(t)\rangle=
\sum_{l=1}^M\langle\sigma(t)|_l\ \hat{h}_{\phi, l}|\sigma^\prime(t)\rangle_l
\prod_{l^\prime\ne l}\delta^{(l^\prime)}_{\sigma\sigma^\prime}.
\end{eqnarray}
This transition matrix, however, is not zero only when the spins of $|\varepsilon_\nu(t)\rangle$ is different 
from those of $|\varepsilon_{\nu^\prime}(t)\rangle$ just on one site 
since the double flip of the spin is absent in the first order perturbation. 
For example, for all $l$'s, the matrix element for the local interaction $\hat{h}_{\phi, l}$ between the state $\nu=1$ and $\nu^\prime=M+2$ vanishes, 
$\langle\varepsilon_1(t)|\hat{h}_{\phi, l}|\varepsilon_{M+2}(t)\rangle=0$.

\bigskip
\bigskip
\bigskip

Let us evaluate the fluctuation by the self-evolution acting 
$\langle\varepsilon_1(t)|$ from the left side of eq. (\ref{expansion-Seq}). 
We obtain the equation for the coefficients, $c_{1N}(t)$ and $c_{2N}(t)$, 
\begin{eqnarray}\label{expansion-1}
\sum_{j=1,2}i\hbar\ \dot{c}_{j1}(t)|\phi_j(t)\rangle
=\sum_{\nu^\prime=1}^N\sum_{j=1, 2}c_{j\nu^\prime}(t)|\phi_j(t)\rangle
\langle\varepsilon_1(t)|\hat{h}_{I}|\varepsilon_{\nu^\prime}(t)\rangle.
\end{eqnarray}
Obviously, at $t=0,$ the matrix element, 
\begin{eqnarray}
\langle\varepsilon_1|\hat{h}_I|\varepsilon_{\nu^\prime}\rangle=\delta_{1, \nu^\prime}\sum_{l=1}^M
\hat{v}_{\uparrow\uparrow}^l,
\end{eqnarray}
is diagonal (i.e., no transition),
while, at $t\ne 0$, the off-diagonal elements exist. 
($\langle\varepsilon_1(t)|\hat{h}_I|\varepsilon_{\nu^\prime}(t)\rangle\ne 0$
for $1\le \nu^\prime\le M+1$.)
We can easily calculate these self-evolution.
For the diagonal element ($\nu^\prime=1$), we obtain, 
\begin{eqnarray}
\langle\varepsilon_1(t)|\hat{h}_I|\varepsilon_1(t)\rangle
&&=
\sum_{l=1}^M\langle\uparrow(t)|\hat{h}_{\phi, l}|\uparrow(t)\rangle_l\\
&&=
\sum_{l=1}^M\hat{v}_{\uparrow\uparrow}^l\cos^2\omega_l t
+\hat{v}_{\downarrow\downarrow}^l\sin^2\omega_l t.
\end{eqnarray}
On the other hand, for the off-diagonal elements, the finite contribution comes from only $M$ states of the total $N$ states, ($2\le \nu^\prime\le M+1$), as we mentioned above:
\begin{eqnarray}
\sum_{\nu^\prime=2}^Nc_{j\nu^\prime}(t)\langle\varepsilon_{1}(t)|\hat{h}_I|\varepsilon_{\nu^\prime}(t)\rangle
&&=\sum_{l=1}^Mc_{jl+1}(t)
\langle\uparrow(t)|\hat{h}_{\phi, l}|\downarrow(t)\rangle_l\\
&&=\sum_{l=1}^Mc_{jl+1}(t)(\hat{v}_{\uparrow\uparrow}^l+\hat{v}_{\downarrow\downarrow}^l)
i\sin\omega_lt\cos\omega_lt.
\end{eqnarray}

At $t=0$, the diagonal element is order of $O(M)$, 
while the off-diagonal elements are exactly zero.
Here, let us assume that the phase ${\omega_l}t$ are distributing randomly around zero at $t\ne 0$.
Then the fluctuation of the on- and off-diagonal elements due to the self-evolution is order of $O(\sqrt{M})$ for sufficiently large $M.$
{\bf Therefore we can neglect the off-diagonal elements compared with the diagonal one 
even taking into account of the fluctuation due to the self-evolution.}

\bigskip

After all, since the total interaction is the sum of the many independent interactions, 
for the states of the environment, 
the transition elements caused by the fluctuation can be ignored compared to the interaction energy (i.e., the diagonal elements) {\bf in statistical sense}.

\bigskip
\bigskip
\underline{\bf Diagonal approximation}
\bigskip
\bigskip

In Zurek's model\cite{Zurek1981}, the scenario that the oscillation in the phase of the state vectors 
selects out the classical states makes use of the random phase mechanism due to 
the interaction energy between the system and its environment. 
In this mechanism, it is important that the phase shift is prior to 
the transition caused by the non-commutativity, $[\hat{h}_I, \hat{h}_0]\ne 0$. 
Namely, the matrix element of the interaction Hamiltonian should be diagonal. 

In the present model, the non-commutativity breaks the diagonality of the system's states, 
but it does not affect on that of the environment's states.
Resultingly, the eigenstate of the environment $|\varepsilon_\nu(t)\rangle$ pins 
the state of the total system, $|\phi\rangle\otimes|\varepsilon\rangle$, 
so that the random phase mechanism can work.

\begin{figure}
\begin{center}
\includegraphics[width=7cm]{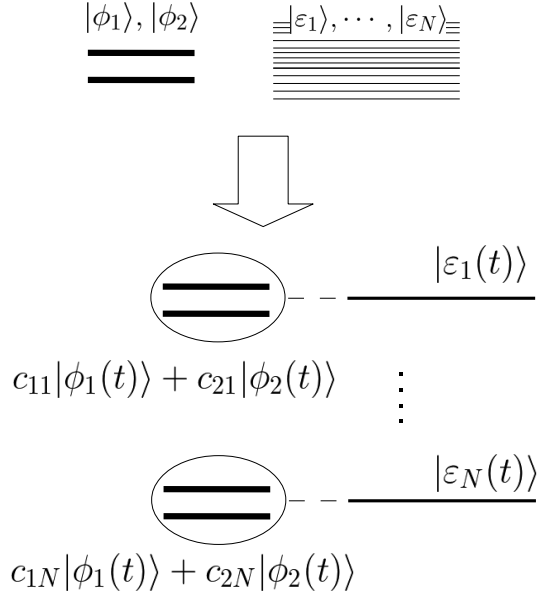}
\caption{The separation to the $N$ states independently time-evoluting each other (image).}
\end{center}
\end{figure}

Next, we evaluate the interaction effect on $\phi.$ 
There are two types of effect, where 
one is the transition $\phi_1\rightleftharpoons\phi_2$ and the other 
is the common phase shift. 
For our purpose, we here take into account only the latter effect by setting  
$c_{i\nu}(t)=\alpha_\nu(t)c_{i\nu}, $ where $|c_{1\nu}|^2+|c_{2\nu}|^2=1.$
For $\alpha_1(t),$ eq. (\ref{expansion-1}) leads, 
\begin{eqnarray}
i\hbar\ \dot{\alpha_1}(t)&&=\alpha_1(t)\lambda_1(t)-
\sum_{l=1}^{M}\alpha_{l+1}(t) \lambda_{l+1, 1}(t)\
i\sin\omega_{l} t\ \cos\omega_{l} t\\
&&\simeq \alpha_1(t)\lambda_1(t).
\end{eqnarray}
As mentioned above, the off-diagonal terms have been neglected for sufficiently large $M$.

\vspace{1cm}

We can re-label the state vectors to stress the `unit' states which are stable against the time-evolution as, 
\begin{eqnarray}
|\nu(t)\rangle:=(c_{1\nu}|\phi_1(t)\rangle+c_{2\nu}|\phi_2 (t)\rangle)|\varepsilon_\nu(t)\rangle, (\nu=1, \dots, N)
\end{eqnarray}
and the above equation becomes, 
\begin{eqnarray}
i\hbar\dot{\alpha}_\nu(t) |\nu(t)\rangle=\hat{h}_I |\nu(t)\rangle.
\end{eqnarray}
The time dependence of $|\nu(t)\rangle$ represents the self-evolution and $\langle\nu(t)|\nu^\prime(t)\rangle=\delta_{\nu\nu^\prime}.$
Then, the perturbation on the phase shift is, 
\begin{eqnarray}
i\hbar\partial_t\alpha_\nu(t)  \simeq \alpha_\nu(t)\lambda_\nu(t),
\end{eqnarray}
where, 
\begin{eqnarray}
\lambda_\nu(t):=\langle \nu(t)|\hat{h}_I|\nu(t)\rangle.
\end{eqnarray}
The coefficient, $\lambda_\nu$,
represents just the interaction energy of the state, $|\nu(t)\rangle$.
We can very easily integrate this and obtain
$\displaystyle \alpha_\nu(t)=\alpha_{\nu}\exp\left[-i\int \lambda_\nu(t) dt/\hbar\right].$
After all, our diagonal approximation leads, 
\begin{eqnarray}
|\Phi(t)\rangle\simeq\sum_{\nu=1}^N\alpha_\nu|\nu(t)\rangle e^{-i\Lambda_\nu(t)/\hbar}, 
\end{eqnarray}
where $\displaystyle\Lambda_\nu(t):=\int\lambda_\nu(t) dt$
represents the time integral of the interaction energy. 

\bigskip

In short, the transition, $|\nu\rangle\to|\nu^\prime\rangle$, is suppressed due to the diagonality of the environment.
So that the state, $|\nu(t)\rangle$, is the minimum unit 
in which the separability of the subsystems is kept. 
Resultingly a kind of localization on $\nu$ enables
the multiple scattering effect on the phase of the state vectors.

\section{Decoherence and Pointer States}

\subsection{Random Phase Mechanism and the Classical State $\nu_c$}

Even for a very weak interaction, $\Lambda_\nu(t)$ is the macroscopic quantity ($O(M)$) and 
occurs very frequent sign inversion on the states, $|\nu(t)\rangle$, with  the index $\nu$ varying.
Therefore, in some cases, 
we can make use of the random phase mechanism, 
where only the states giving the extreme values to $\Lambda_\nu(t)$ can survive as,
\begin{eqnarray}\label{rpa}
|\Phi(t)\rangle\simeq\sum_{\nu_c}\tilde{\alpha}_{\nu_c}|\nu_c(t)\rangle e^{-i\Lambda_{\nu_c}(t)/\hbar}.
\end{eqnarray}
Here $\nu_c$'s satisfy the condition, $\delta \Lambda_\nu(t)/\delta\nu =0.$\footnote{
In general, since $\nu_c$'s change according to time, the time dependence of 
$\Lambda_{\nu_c}(t)$ will draw the envelope of $\{\Lambda_\nu(t)\}$.}

The transition due to general perturbation is such the case. 
When we choose a specific state vector to follow its time evolution, 
the state, $|\nu_c(t)\rangle$, has a special meaning. 
Although, as confirmed above, every member of $\{\nu\}$ is closed against the interaction causing the decoherence,
almost all of these states are not stable against various other interaction at all.
Only $|\nu_c(t)\rangle$'s are stable in such the transition between these states, since the destructive interference occurs in the transition matrix as,
\begin{eqnarray}
\sum_{\nu^\prime}\alpha_\nu\langle\nu^\prime(t)|\hat{H}_I|\nu(t)\rangle e^{i(\Lambda_{\nu^\prime}-\Lambda_\nu)/\hbar}\simeq
\sum_{\nu_c^\prime}\alpha_\nu\langle\nu_c^\prime(t)|\hat{H}_I|\nu(t)\rangle e^{i(\Lambda_{\nu_c^\prime}-\Lambda_\nu)/\hbar}.
\end{eqnarray}

We here notice that, 
the phase factor, $\Lambda$, does not affect the physical quantities of the total state.
We see this in \S 4.3.
It is no surprising 
that, for a weak interaction, 
the system's density matrix cannot approach to 
the approximate mixture of states.

\bigskip
\bigskip
\underline{\bf Classicality}
\bigskip
\bigskip

In the state, $|\nu(t)\rangle$, 
the state-vector of the subsystem, $c_{1\nu}|\phi_1(t)\rangle+c_{2\nu}|\phi_2(t)\rangle$, have its own Schr\"odinger equation
with the `external' potential, $\hat{v}:=\langle\varepsilon_{\nu}|\hat{h}_I|\varepsilon_{\nu}\rangle$.
In other words, the subsystem $\phi$'s are {\it separable} 
in each branch, $|\nu(t)\rangle$.
In these states, only the states, $|\nu_c(t)\rangle$'s, are stable for 
various perturbative interactions but the others are not:
the unstable states, $\nu\ne\nu_c$, are without separability and have transition elements 
for such the interactions as, 
$\langle \nu(t)|\hat{H}_I|\nu^\prime(t)\rangle$, each other.

These features (having the separability and refusing the superposition) are what we expect 
a classical system to have.
For example, in our daily experience, 
the COM degree of freedom of a macro-object is localized 
at the same time that it has a equation of motion of itself.

\bigskip

The density matrix corresponding to the approximate mixture of states 
is also the small part of the total density matrix 
when the self-Hamiltonian is dominant (Fig. 4): 
\begin{eqnarray}
|\Phi(t)\rangle\langle\Phi(t)|&&:=\sum_{\nu} p_{\nu}|\nu(t)\rangle\langle\nu(t)|\\
&&=|\Phi_c(t)\rangle\langle\Phi_c(t)|+\sum_{\nu\ne\nu_c} 
p_{\nu}|\nu(t)\rangle\langle\nu(t)|, \\
|\Phi_c(t)\rangle\langle\Phi_c(t)|&&:=\sum_{\nu_c} p_{\nu_c}
|\nu_c(t)\rangle\langle\nu_c(t)|.
\end{eqnarray}
We note that the above random-phase mechanism makes use of the entanglement between a system and its environment.
On the other hand, if we start from the product state, the initial state is,
\begin{eqnarray}
|\Phi(0)\rangle=(c_1|\phi_1\rangle+c_2|\phi_2\rangle)(\alpha_1|\varepsilon_1\rangle+ \cdots+\alpha_N|\varepsilon_N\rangle)
=\sum_{\nu=1}^N\alpha_\nu(c_1|\phi_1\rangle+c_2|\phi_2\rangle)| \varepsilon_\nu\rangle.
\end{eqnarray}
Here each state has same superposition state of the subsystem $\phi$
so that the random-phase mechanism does not work.
It, however, is natural that the various interaction in a macro-object 
makes the coefficients $\{c_i\}$ depend on $\nu$.

\subsection{Example: Localization of the Center-of-mass}

The early studies (for example, see \cite{Joos2003}) demonstrate 
the decoherence between the localized states of a 
macro-object using the S-matrix of the scattering by the environment.
Although our present approximation neglects such the transition,  
we can show the stability and the emergence of the localization  
in the state vector.

For a short-time perturbation, the approximation,
$\Lambda_\nu(t)\simeq \lambda_\nu\Delta t$, leads 
that the surviving state $|\nu_c(t)\rangle$ is the approximate eigenstate of the interaction Hamiltonian.
Moreover, the state $\phi$ being at the center of the interaction remarkably 
affects $\Lambda_\nu$: in many cases, 
we can assume that the system's state $\phi$ is in the `external' potential, $\hat{v}:=\langle\varepsilon_\nu|\hat{h}_I|\varepsilon_\nu\rangle$, being independent of $\nu$. 
This assumption does not affect the previous section's consequence.
Therefore the interaction energy, $\lambda_\nu$, becomes to depend only on the state of the system, 
$c_{1\nu}|\phi_1(t)\rangle+c_{2\nu}|\phi_2(t)\rangle$.
Especially, 
when the states, $|\phi_1\rangle$ and $|\phi_2\rangle$, correspond to the position in space 
(namely $|\phi_i\rangle=|{\bf R}_i\rangle$), 
this leads the localization of the COM (Fig. 5). 
We here notice that the total state is still in the superposition of states. 

\begin{figure}
\begin{center}
\includegraphics[width=7cm]{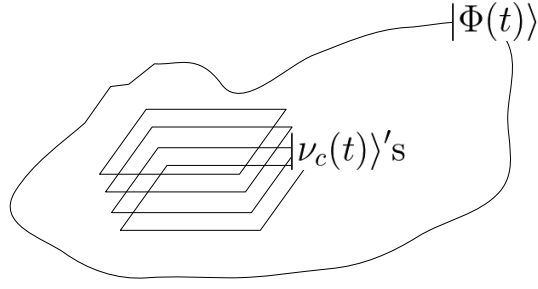}
\caption{The pointer states, $|\nu_c(t)\rangle$'s, in the total state, $|\Phi(t)\rangle$ (image).}
\end{center}
\end{figure}

\subsection{Energy Conservation}

The time-dependent state vector, 
\begin{eqnarray}
|\nu(t)\rangle=e^{-i\hat{h}_0t/\hbar}|\phi_\nu\rangle e^{-i\hat{h}_\varepsilon/\hbar}|\varepsilon_\nu\rangle,
\end{eqnarray}
represents the time evolution by the self-Hamiltonian $\hat{h}_0$ and $\hat{h}_\varepsilon$.
Including the effect of the lowest order of the interaction, 
the state of the total system is, 
\begin{eqnarray}
|\Phi(t)\rangle=\sum_{\nu}\alpha_\nu|\nu(t)\rangle e^{-i\Lambda_\nu(t)/\hbar}, 
\end{eqnarray}
where $\displaystyle\Lambda_\nu(t)=\int\langle\nu(t)|\hat{h}_I|\nu(t)\rangle dt$ is the time integrated interaction energy of 
the state $\nu$.
Within our lowest order approach, only the phase factors have changed by the interaction so that the physical quantities are not affected:
in fact, the orthogonality relation of the environmental state, 
$\langle\varepsilon_\nu|\varepsilon_{\nu^\prime}\rangle=\delta_{\nu\nu^\prime}$, leads, 
\begin{eqnarray}
\langle\Phi(t)|\hat{h}_0|\Phi(t)\rangle=\sum_{\nu}|\alpha_\nu|^2\langle\nu(t)|\hat{h}_0|\nu(t)\rangle\\
=\sum_{\nu}|\alpha_\nu|^2\langle\phi_\nu(t)|\hat{h}_0|\phi_\nu(t)\rangle\\
=\langle\phi(t)|\hat{h}_0|\phi(t)\rangle=E_0.
\end{eqnarray}
We here notice that not only the classical states ($\nu=\nu_c$) but also the quantum states ($\nu\ne\nu_c$) 
equally contribute to the energy. 
There is no influence of the decoherence.

\begin{figure}
\begin{center}
\includegraphics[width=7cm]{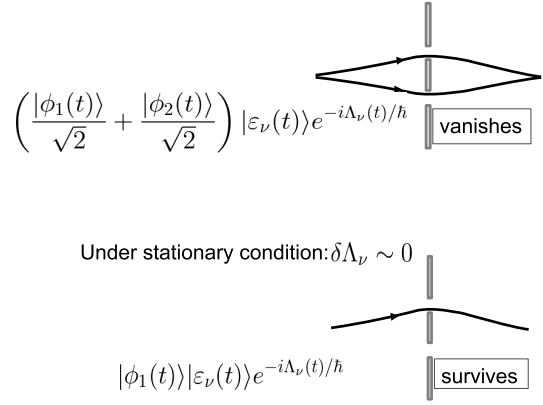}
\caption{The localized state corresponding to the stationary phase (image).}
\end{center}
\end{figure}

%\newpage

\section{Decoherence and Order Formation}

We have found out the decoherence subspace, 
i. e., $|\nu_c(t)\rangle$.
While the stable state, $|\phi_{\nu_c}(t)\rangle$, of the subsystem 
is separable in it,  
it is possible to have variety of features originating from the self-Hamiltonian in a weak-interaction case. 

The localization of the COM's state, $|\phi_{\nu_c}(t)\rangle$, 
makes the deep valley in the potential 
in contrast to other super position of states. 

For example, in the branch rabeled by $\nu_c$, composed of two macroscopic objects with 
electric charges of the opposite sign, 
the rapid combining of these objects will be possible.
The potential structure with the deep valley also enables the resulting states to be energetically stable.

Although this process is still a unitary time evolution, 
this branch appears to be
accompanied with the dissipation.
The separability, i.e., the separation of 
macro- and microscopic degrees of freedom, is probably 
relevant to the time asymmetric appearance of our dairy experience. 
The former, $|\phi_{\nu_c}\rangle$, deserves the minimum unit 
composing macroscopic order due to its separability 
while the whole branch, $|\nu_c(t)\rangle$, goes to the 
energitically uniform in space by the unitary evolution. 
von Neumann's paper\cite{Neumann2010} gives the proof to 
the dissipation due to the quantum unitarity 
starting from such the special initial condition.

However, the inconsistency in appearance between 
the arrow of time and the quantum unitarity is 
a wide and complicated topic. 
For example, the role of consciousness may be relevant. 
More sophisticated discussion is given by Zeh\cite{Zeh2009}.

\section{Conclusions}

We investigated how the scenario of the quantum decoherence 
is justified 
even when the system-environment interaction is weak:
in this case, from the perspective of the energy conservation, 
the contribution of the off-diagonal terms are obviously important. 
On the two state system weakly interacting with its environment, 
we find that the classical states, i.e. the pointer states, 
surely exist but constitute only the small part of a total state vector. 
(Only the degrees of freedom being at the center of interaction 
can has the pointer states. )
This point escapes our notice if we only study strong interaction cases.

The existential criterion given by Zurek\cite{Zurek2007} may be still useful 
for our weak interaction case: 
our study indicates that, the normal quantum state, $|\nu\rangle$, 
exists but the pointer state, $|\nu_c\rangle$, exists objectively. 
(Also the former state contributes to the energy conservation in the whole 
unitary evolution.)

We have made use of the effect of the decoherence in 
the transition matrix so that 
the classical states are stable against not only the system-environment interaction 
but also general perturbation.

For various realistic situations, to study 
under what conditions  
the quantum-classical boundary emerges 
is practically important. 
For example, 
the concept of quantum transition is often 
used empirically: the chemical reaction
is usually recognized to be a probabilistic process.
The important relation between the quantum decoherence and various fields 
has been already discussed by Joos and Zeh\cite{Joos1985}.

\end{document}